\documentclass{article} 
\usepackage[preprint]{colm2026_conference}

\usepackage{microtype}
\usepackage{hyperref}
\usepackage{url}
\usepackage{booktabs}
\usepackage{graphicx}
\usepackage{amsmath}
\usepackage{enumitem}
\usepackage{lineno}
\usepackage{subcaption}
\usepackage{lipsum}
\usepackage{csquotes}
\usepackage{algorithm}
\usepackage{algorithmic}
\usepackage{wrapfig}
 \usepackage{tcolorbox}

 \usepackage[T1]{fontenc}
\usepackage[utf8]{inputenc} 

\definecolor{darkblue}{rgb}{0, 0, 0.5}
\hypersetup{colorlinks=true, citecolor=darkblue, linkcolor=darkblue, urlcolor=darkblue}

\title{Planning to Explore: \\
Curiosity-Driven Planning for LLM Test Generation}

\author{Alfonso Amayuelas$^1$\thanks{Work partially completed during a visit to King Abdullah University of Science and Technology}
, Firas Laakom$^2$, Piotr Pi\k{e}kos$^2$, Wenyi Wang$^2$,\\
\textbf{Yifan Xu$^2$, Yuhui Wang$^2$, Jürgen Schmidhuber$^2$, William Wang$^1$}\\
$^1$University of California, Santa Barbara\\
$^2$King Abdullah University of Science and Technology\\
\texttt{amayuelas@ucsb.edu, \{firas.laakom, piotr.piekos, wenyi.wang,}\\
\texttt{yifan.xu, yuhui.wang, juergen.schmidhuber\}@kaust.edu.sa}, \texttt{william@cs.ucsb.edu} \\
}

\begin{document}

\ifcolmsubmission
\linenumbers
\fi

\maketitle

\begin{abstract}
The use of LLMs for code generation has naturally extended to code testing and evaluation. As codebases grow in size and complexity, so does the need for automated test generation. Current approaches for LLM-based test generation rely on strategies that maximize immediate coverage gain, a greedy approach that plateaus on code where reaching deep branches requires setup steps that individually yield zero new coverage. Drawing on principles of Bayesian exploration, we treat the program's branch structure as an unknown environment, and an evolving coverage map as a proxy probabilistic posterior representing what the LLM has discovered so far. Our method, \emph{CovQValue}, feeds the coverage map back to the LLM, generates diverse candidate plans in parallel, and selects the most informative plan by LLM-estimated Q-values, seeking actions that balance immediate branch discovery with future reachability. Our method outperforms greedy selection on TestGenEval Lite, achieving 51--77\% higher branch coverage across three popular LLMs and winning on 77--84\% of targets.
In addition, we build a benchmark for iterative test generation, \emph{RepoExploreBench}, where they achieve 40-74\%. These results show the potential of curiosity-driven planning methods for LLM-based exploration, enabling more effective discovery of program behavior through sequential interaction.
\end{abstract}

\begin{figure}[h]
    \centering
    \includegraphics[width=.85\textwidth]{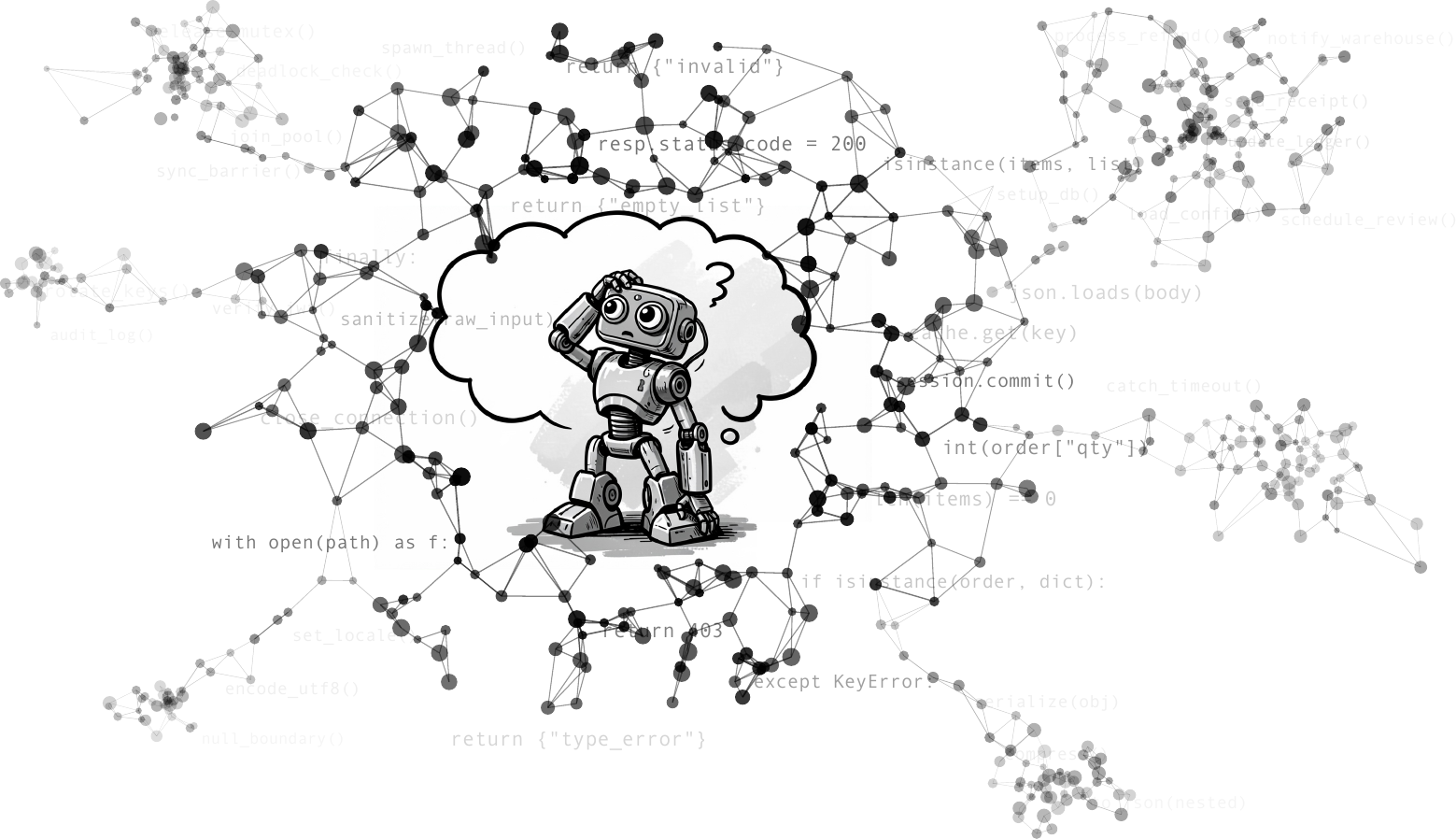}
    \label{fig:placeholder}
\end{figure}

\section{Introduction}
\label{Sec:Introduction}

LLMs are increasingly used not only to generate code, but also to test it \citep{chen2021evaluating,schafer2023empirical}. However, generating tests that reach deep program behavior often requires discovering the right sequence of steps and branches, where some steps can yield zero immediate coverage gain. Most existing LLM-based test generation methods are effectively greedy. They generate tests independently, without learning from previous executions or reasoning about what remains unexplored. This challenge aligns with a well-studied problem in reinforcement learning where an agent needs to explore an unknown environment efficiently. We apply this perspective to LLM-based test generation. We view iterative test generation as the exploration of an unknown program, where each executed test reveals part of the program's behavior through the branches it reaches. Rather than treating coverage solely as an evaluation metric, we feed back a structured coverage map to the LLM and use it as the basis for information-gain-driven action selection.



For decades, researchers have been exploring how to transform curiosity into a computational feature that can be integrated into modern AI systems \citep{schmidhuber2010formal}. In exploration theory, \citet{schmidhuber1991possibility, schmidhuber2010formal} formalized curiosity as the drive to maximize learning progress by seeking experiences that improve the agent's world model. \citet{sun2011planning} proved that, in the Bayesian setting, the optimal strategy for maximizing cumulative information gain requires planning ahead, not greedy selection, because information gain is additive only in expectation. They demonstrated this with the corridor problem, where a deterministic passage connects two information-rich regions. Greedy exploration ignores the corridor (zero immediate information gain); only planning-aware exploration traverses it. Although our setting uses branch count as a proxy for mutual 
information, the core insight that greedy selection blocks 
future discovery applies to our setting.

With the expansion of LLMs, we have also seen an increase in their use to generate code. And therefore, the interest for LLM-based test generation. This is the case where an LLM needs to interact iteratively and discover an unknown environment. Tools like CodaMosa \citep{lemieux2023codamosa}, CoverUp \citep{altmayer2025coverup}, TestForge \citep{jain2025testforge} generate tests that improve code coverage. But they all share a fundamental limitation, they are greedy. They target specific coverage gap they see right now. When reaching deep branches requires a sequence of setup steps that individually contribute zero coverage ( e.g. import chains, class initialization, passing validation gates), greedy strategies plateau and they do not cover all the code. 

In this work, we formalize LLM-based test generation as Bayesian exploration of an unknown environment \citep{sun2011planning}. In this environment, the program's branch structure is the unknown parameter, the coverage map summarizes the agent's exploration state, and the LLM with its interaction history serves as the world model. Our method, \emph{CovQValue}, generates various candidate test plans, then scores each one using an LLM-estimated curiosity Q-value that aims to balance immediate branch discovery and future reachability. After, it executes the highest-scoring plan and loops again. We conduct experiments on TestGenEval Lite, an existing benchmark built on SWE-Bench; and on \emph{RepoExploreBench}, a benchmark we introduce for iterative test generation, built on 93 modules from 9 popular Python packages. We run the experiments on three popular LLMs (Gemini 3 Flash, GPT-5.4-mini and Mistral Large 3). CovQValue outperforms greedy selection on RepoExploreBench by 40-74\% more branch coverage, and on TestGenEval Lite by 51--77\%, winning on 77--84\% of targets, with the largest gains on repositories with deep corridor structure.

Our contributions can be summarized as follows:
\begin{enumerate}[noitemsep,leftmargin=*]
    \item \textbf{We formalize LLM test generation as Bayesian Exploration}, where the coverage map serves as the sufficient statistic summarizing the agent's exploration state, and the iterative feedback loop is the core mechanism. This provides a theoretically grounded alternative to greedy coverage-directed approaches.
    
    \item \textbf{We release \underline{RepoExploreBench}},  a benchmark of 93 modules from 9 popular Python packages designed to evaluate iterative, exploration-based test generation. Existing benchmarks evaluate one-shot generation; RepoExploreBench targets code with corridor structure where multi-round feedback is essential.
    
    \item \textbf{We introduce \underline{CovQValue}, a method for iterative code test generation} that combines coverage map feedback, diverse plan generation, and Q-value plan selection to maximize branch discovery. Evaluated on our benchmark, RepoExploreBench, and well-known benchmark, TestGenEval, with 3 different LLMs, CovQvalue shows better performance than greedy selection by 40--77\% on branch coverage across both benchmarks more, winning on 77--84\% of targets.
\end{enumerate}

Beyond test generation, planning-aware exploration can apply to other agents that interact sequentially with an unknown environment (e.g. API discovery, scientific experimentation). Our results suggest that the gap between greedy and exploratory strategies widens when the environment contains corridor structure, a property common to many real-world domains. 

\vspace{-3pt}
\section{Related Work }
\label{Sec:Related_Work}
\vspace{-3pt}

\paragraph{Artificial Curiosity and Optimal Exploration} Curiosity is commonly defined as the motivation that drives exploration and problem-solving in artificial agents \citep{singh2010intrinsically}. \citet{schmidhuber1991possibility} describes intrinsic motivation as the drive to improve one’s predictive world model. Later, \citet{schmidhuber1991curious} expands this theory into a general framework that formalizes curiosity as an optimization problem in reinforcement learning, where the goal is to maximize the improvement of a predictive world model. \citet{storck1995reinforcement} instantiated this principle by using KL divergence between successive probability estimates as the curiosity reward in non-deterministic environments. Building on this, \citet{sun2011planning} to show theoretically how an agent can optimally plan action sequences based on previous experiences such that the cumulative expected information gain is maximized.

\paragraph{LLM-based Test Generation} Traditionally, automated test generation has relied on test-based software tests, where tools like EvoSuite \citep{fraser2011evosuite} or Pynguin \citep{lukasczyk2022pynguin} to mutate test inputs and maximize coverage. These tools operated at the input level without generating semantic test scripts. LLMs and their use for writing code has increased tremendously due to their good results and the verifiable nature of code \citep{chen2021evaluating}. The rise of automatically generated code has increased the  need for automated testing as a self-verification mechanism. ChatUniTest \citep{chen2024chatunitest} or ChatTester \citep{yuan2024evaluating} are examples of this. LLMs can generate tests, but coverage plateaus on complex code. \citet{jain2024testgeneval} introduces TestGenEval to measure this. It is a file-level test generation benchmark on real-world projects built on SWE-Bench \citep{jimenez2023swe}. Existing approaches for LLM-based test generation are single-step or reactive. CodaMosa \citep{lemieux2023codamosa} presents a hybrid method that detects coverage stalls and calls the LLM reactively. CoverUp \citep{altmayer2025coverup}, TELPA \citep{yang2024advancing} and \citet{xu2026enhancing} present solutions that treat coverage as a sequence of independent single-step optimization problems, where test at $n$ does not affect $n+1$. Therefore, maximizing immediate coverage gain. They rely on methods that generate tests, measure coverage, feed the uncovered branches and generate more tests. TestForge \citep{jain2025testforge} takes an agentic approach, iteratively refining tests based on execution feedback, but still selects actions greedily. None of these approaches reason about how covering intermediate branches opens access to deeper ones.

\paragraph{Information Gain to guide LLM Actions} A recent line of work tries to apply information-gain maximization to guide LLM behavior at inference time. Uncertainty of Thoughts \citep{hu2024uncertainty} uses tree search with information-gain rewards to select questions that reduce the LLMs uncertainty. BED-LLM \citep{choudhury2025bed} formalizes this more rigorously as sequential Bayesian Experimental Design (BED), creating estimators to maximize the expected information gain derived from the LLM's predictive distribution. CuriosiTree \citep{cooper2025curious} takes a similar heuristic approach to zero-shot information acquisition. These methods operate in static environments where the questions do not change the state of the world. Test generation introduces a different challenge where, although the program's branch structure is static, the agent's knowledge of how to reach deep branches evolves with each execution. Reaching certain branches requires discovering specific setup sequences through prior tests, creating epistemic corridors analogous to those in \citet{sun2011planning}'s MDP experiments, where only multi-step planning succeeds.

\section{Method}
\label{Sec:Method}

\paragraph{Test Generation inspired by Bayesian Exploration}
We frame iterative test generation as a heuristic approximation of Bayesian exploration in an unknown environment \citep{sun2011planning}. The agent interacts with a program whose internal branch reachability is initially unknown. At each step, the agent generates a test (action), executes it, and observes which branches are reached (observation). The goal is to choose tests such that cumulative branch coverage (our measure of knowledge about the program) grows as quickly as possible.

\begin{table}[b]
\centering
\resizebox{\textwidth}{!}{%
\begin{tabular}{@{}l l l@{}}
\toprule
\textbf{Definitions} & \textbf{Symbol} & \textbf{Test Generation} \\
\midrule
Unknown environment & $\Theta$ & Program's branch reachability structure \\
Prior & $p(\Theta)$ & No branches observed \\
History & $h = (a_1,o_1) \cdots (a_t,o_t)$ & Sequence of (test, coverage result) pairs \\
Exploration state & $p(\Theta | h)$ & Coverage map $\mathcal{C}_t$ (sufficient statistic after $t$ executions) \\
Action & $a$ & Test plan (sequence of $S$ scripts) \\
Observation & $o$ & Branches hit by executing the plan \\
Expected info gain & $\bar{g}(a|h) = I(O; \Theta \mid h, a)$ & Expected new branches from plan $a$ \\
Curiosity Q-value & $q(a|h) = \Tilde{g}(a|h)+ \gamma \cdot \mathbb{E}_{o|h,a}[v(hao)]$ & Immediate gain $+$ future reachable branches \\
\midrule
Corridor & & Validation gates, setup sequences \\
\bottomrule
\end{tabular}}%
\caption{Formal mapping from \citet{sun2011planning} optimal Bayesian exploration framework to LLM-based test generation.}
\label{tab:framework_mapping}
\end{table}

This is grounded in the theory of \citet{sun2011planning}, which presents a study of optimal exploration in dynamic environments. An agent interacts with an environment in discrete time steps, performing actions and receiving observations. The environment is characterized by an unknown parameter $\Theta$, and the agent maintains a posterior $p ( \Theta | h ) $ over $\Theta$ given the interaction history $h = (a_1, o_1), ...., (a_t, o_t)$. The expected information gain of performing action, $a$, given history, $h$, is:

\begin{equation}
\label{equation:g_hat}
   \Tilde{g}(a \mid h) = I(O; \Theta \mid h, a)
\end{equation}

$I(\cdot;\cdot)$ is the mutual information between the unknown parameter, $\Theta$ and the observation, $O$, conditioned on history, and action. Because information gain is additive only in expectation \citet{sun2011planning}, it cannot be directly substituted as a reward in standard reinforcement learning. The optimal exploration policy requires a recursively defined curiosity Q-value that yields an optimal exploration policy:

\begin{equation}
\label{equation:q_value}
    q(h, a) = \Tilde{g}(a \mid h) + \gamma \cdot \mathbb{E}_{o \mid h,a}[v(hao)]
\end{equation}

where $v(h) = \max_a q(h, a)$ is the curiosity value of history $h$, and $\gamma$ is a discount factor. The Q-value captures not just the immediate expected information gain of action $a$, but also the expected value of the state the agent ends up in after taking $a$ and observing $o$. Table \ref{tab:framework_mapping} shows how we instantiate this framework for test generation. We represent our framework

\paragraph{Coverage Maps as Exploration State} One challenge in applying the framework with LLMs is representing the agent's epistemic state, analogous to the posterior $p(\Theta|h)$. Proxy signals such as sampling entropy and multi-model disagreement are poorly calibrated for epistemic uncertainty about program behavior. We avoid this problem by using the coverage map, the set of branches observed across all test executions, as the sufficient statistic summarizing the agent's exploration state. In the original framework, the unknown parameter $\Theta$ corresponds to the program's full branch reachability structure, and each newly discovered branch directly reduces uncertainty about $\Theta$. The coverage map does not encode a full probability distribution over unseen branches because it is computationally intractable. Rather, it serves as a deterministic proxy of the posterior that the LLM conditions on to perform implicit probabilistic reasoning about which branches remain reachable. The coverage map updates after each execution ($\mathcal{C}_t = \mathcal{C}_{t-1} \cup B(a_t)$) and is then fed back to the LLM as structured text, enabling it to condition generation on the current state of exploration.

\begin{figure}[t]
    \centering
    \vspace{-14pt}
    \includegraphics[width=.98\textwidth]{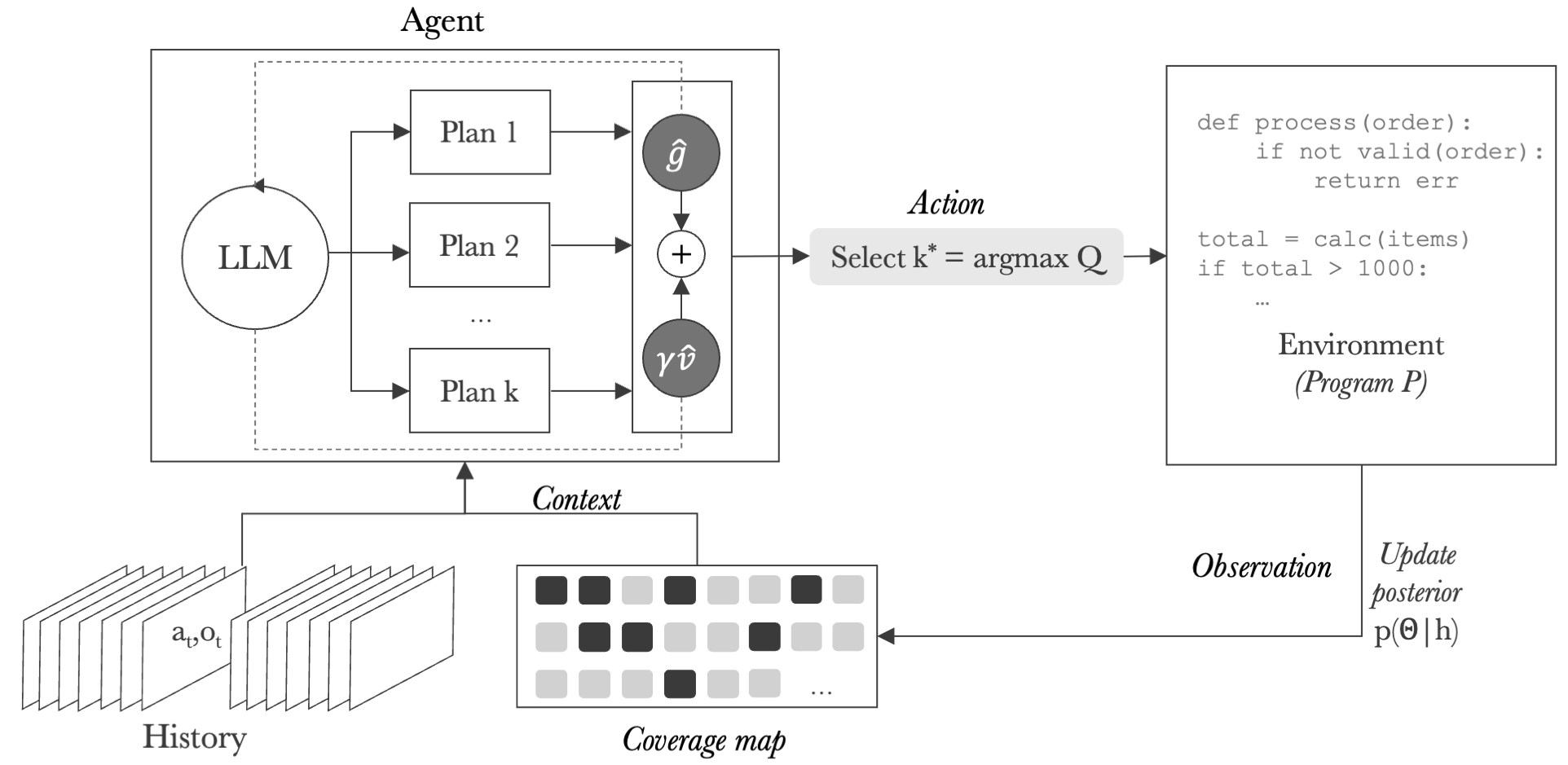} 
    \caption{\textbf{Overview of CovQvalue}. At each round, the LLM generates $K$ diverse candidate plans conditioned on the coverage map. Each plan is scored by its estimated Q-value, information gain plus discounted future reachability (Equation \ref{equation:q_value}), and the highest scoring plan is executed. The resulting branch coverage updates the coverage map for next round.}
    \vspace{-10pt}
    \label{fig:framework_method}
\end{figure}


\paragraph{Curiosity Q-values for Plan Selection} In theory, an optimal exploration policy selects the action maximizing the curiosity Q-value $q(h, a)$. Computing this exactly is intractable, as it is exponential in the planning horizon. Thus, We approximate it using the LLM itself as a heuristic estimator. While this does not yield a mathematically rigorous value function, it leverages the LLM's internal coding patterns to recognize which test sequences are likely to unlock deep code paths. To generate diverse candidates, each of the $K$ plans is prompted with a different exploration directive. For example, targeting main functionality, error-handling paths, or cross-module interactions. This encourages the LLM to propose structurally different test sequences rather than $K$ variations of the same approach. 

Given a candidate test plan and the current coverage map, the LLM estimates two quantities: (1) the immediate expected gain $\hat{g}$ (Equation \ref{equation:g_hat}), which represent how many new branches the plan is likely to discover; and (2) the future reachability $E[v(h')]$, how many additional branches will become reachable as a consequence of executing the plan, for instance by passing validation gates or establishing setup sequences. The curiosity Q-value is then the Q-value in Equation \ref{equation:q_value}, where $\gamma$ balances the two terms. Notably, the method only requires the ranking of plans, not precise Q-value estimates. In this open-loop, each plan is scored and committed to without intermediate feedback, a practical approximation that amortizes scoring cost across $S$ executions while enabling coherent multi-step test sequences.

In practice, we prompt the LLM to predict two values on a 0–50 scale: (1) the expected new branches ($\hat{g}$) and (2) additional branches made reachable for future tests ($\hat{v}$). The Q-value is calculated as $Q = \hat{g} + \gamma \cdot \hat{v}$. For example, if the LLM outputs $(12, 20)$ and $\gamma = 0.5$, the Q-value would be $12 + 0.5 \times 20 = 22$. We then select the plan with the highest Q-value from $K$ candidates. Since only the relative ranking matters, the absolute scale of predictions does not affect the selection, as long as the ordering stays consistent. Full prompt templates are available in Appendix~\ref{app:prompts}.

\paragraph{Coverage Corridors} The advantage of planning with Q-values may be more pronounced on code with what we call \textit{corridor} structure. These corridors are sequences of setup steps that each contribute zero new branch coverage individually, but collectively unlock access to deep logic. Programs with input validation gates, initialization sequences, or import-chain dependencies contain such corridors. Greedy strategies plateau on such code because every candidate that passes setup appears equally uninformative. The Q-value scorer resolves this by recognizing that plans investing in setup have high future reachability, selecting them over plans that target surface-level branches. This follows from the recursive structure of the Q-value (Equation~\ref{equation:q_value}), where the future-value term $\mathbb{E}[v(hao)]$ assigns positive value to actions with zero immediate gain but high possible reachability.

\section{Experiments}
\label{Sec:Experiments}

\paragraph{Benchmarks} We evaluate whether coverage-map-guided planning improves branch coverage over standard baselines across diverse real-world code. To validate our methods, we use two benchmarks including 233 real-world Python source files across 19 repositories:

\textbullet\hspace{0.5em} \textbf{TestGenEval Lite} \citep{jain2024testgeneval} is an existing benchmark of 160 file level test generation targets from SWE-bench \citep{jimenez2023swe}. It is built over 11 Python repositories including \texttt{Django, SymPy, pytest, and matplotlib}. We evaluate on 140 files, excluding scikit-learn which requires per container C extension compilation. Each target runs in a repository-specific SWE-bench Docker testbed with the correct Python environment and version-pinned dependencies. 

Since TestGenEval Lite and other benchmarks evaluate one-shot generation rather than iterative exploration, we introduce RepoExploreBench for the 
iterative setting. In our experiments, we report results consistent in both benchmarks.

\textbullet\hspace{0.5em} \textbf{RepoExploreBench} (Ours) comprises 93 modules from 9 popular open-source Python packages: \texttt{click, requests, flask, rich, jinja2, httpx, pydantic, werkzeug, and starlette}. Packages were selected from the top 150 most-downloaded PyPI packages using three criteria: (1) Python source with $\geq$5K lines, (2) no C extensions or build tools, and (3) exhibit corridor structure (e.g. import chains, class hierarchies, and configuration requirements). From each package, we select modules with $\geq$200 lines, yielding 93 targets totaling $\sim$77K lines of code. All modules execute inside a Docker container with packages pre-installed. We provide more details about the benchmark in Appendix \ref{app:RepoExploreBench}.

\paragraph{Strategies} We compare four strategies, all receiving the same execution budget:

  \begin{itemize}[leftmargin=*]
  
    \item \textbf{Random}: generates $K{=}3$ test scripts from a standard prompt containing the module source code and recent test history. One script is selected uniformly at random.
    \item \textbf{Greedy}: same generation as Random, but the LLM selects which candidate is ``most likely to cover new code paths.'' This represents standard practice in coverage-directed generation \citep{altmayer2025coverup}.
    \item \textbf{CovGreedy}: augments the generation prompt with the coverage map $\mathcal{C}_t$. The coverage map is a structured text block reporting branches discovered, recent discovery rate, stagnation warnings, and the most informative prior tests. Its full format and evolution are shown in Appendix~\ref{app:coverage_map}. The LLM is instructed to target undiscovered branches. Selection is random from $K{=}3$ candidates, differing in this point from the Greedy Selection mechanism. The component ablation in Section~\ref{Sec:Results} isolates these factors.
    \item \textbf{CovQValue} (ours): generates $K{=}3$ diverse trajectory plans, each consisting of $S{=}3$ sequential test scripts. Each plan receives a different diversity hint (e.g., main functionality, error handling, interactions). Plans are scored by the LLM-estimated Q-value from Equation~\ref{equation:q_value} with $\gamma{=}0.5$, and the highest-scoring plan is executed.
  \end{itemize}

\paragraph{Models} We evaluate with three LLMs to test generalization across model families: Gemini 3 Flash (Google), GPT-5.4 Mini (OpenAI), and Mistral Large 3 (Mistral AI). The same model is used for all components within a run: test generation, Q-value scoring, and greedy selection. Generation uses temperature 0.9; scoring and selection use temperature 0.3.

\paragraph{Metrics} To measure the effect of our methods, we use \emph{branch coverage}, which is the cumulative number of unique branches executed across all test scripts. Branch coverage directly measures how much of the program's behavior has been discovered, making it the natural metric for an exploration framework. Additionally, we also report \textit{line coverage} to confirm that results hold under the metric used by TestGenEval \citep{jain2024testgeneval}, and \textit{pass rate} (fraction of generated scripts that execute without error) to differentiate whether coverage gains come from writing more valid tests or from targeting more informative code paths. Both are reported in Appendix~\ref{app:lines}.

\paragraph{Execution Budget} All strategies receive exactly the same number of test script executions ($N{=}24$). Random, Greedy, and CovGreedy each execute 24 individual scripts. CovQValue executes 8 plan rounds of 3 scripts each ($8 \times 3 = 24$),  updating the coverage map once per round (8 updates total). CovQValue additionally requires $K{=}3$ Q-value scoring calls per round. Thus, while total executions are equalized, the strategies differ in feedback frequency and total LLM calls. Despite receiving fewer feedback updates, CovQValue outperforms baselines, suggesting that multi-step planning compensates for less frequent posterior updates.

\section{Results}
\label{Sec:Results}


\begin{table*}[t]
\centering
\resizebox{0.98\textwidth}{!}{
\begin{tabular}{l ccc ccc}
\toprule
& \multicolumn{3}{c}{\textbf{RepoExploreBench}}& \multicolumn{3}{c}{\textbf{TestGenEval Lite}} \\
\cmidrule(lr){2-4} \cmidrule(lr){5-7}
& Gemini 3 & GPT-5.4 & Mistral & Gemini 3 & GPT-5.4 & Mistral \\
& Flash & mini & Large 3 & Flash & mini & Large 3 \\
\midrule
Random & 43.2 {\tiny$\pm$4.5} & 31.2 {\tiny$\pm$3.0} & 41.2 {\tiny$\pm$4.2} & 64.4 {\tiny$\pm$6.3} & 50.4 {\tiny$\pm$4.6} & 65.9 {\tiny$\pm$5.9} \\
Greedy & 41.8 {\tiny$\pm$3.9} & 28.6 {\tiny$\pm$2.8} & 42.3 {\tiny$\pm$4.4} & 61.5 {\tiny$\pm$6.0} & 53.2 {\tiny$\pm$4.8} & 65.0 {\tiny$\pm$5.8} \\
CovGreedy & 51.7 {\tiny$\pm$5.7} & 36.5 {\tiny$\pm$3.5} & 40.0 {\tiny$\pm$4.5} & 77.6 {\tiny$\pm$7.3} & 66.0 {\tiny$\pm$6.2} & 74.7 {\tiny$\pm$7.1} \\
CovQValue & \textbf{72.4} {\tiny$\pm$6.5} & \textbf{49.8} {\tiny$\pm$5.0} & \textbf{59.1} {\tiny$\pm$6.0} & \textbf{108.9} {\tiny$\pm$9.0} & \textbf{80.5} {\tiny$\pm$7.2} & \textbf{100.0} {\tiny$\pm$8.4} \\
\midrule
$\Delta$ vs Greedy & +30.6 & +21.2 & +16.8 & +47.4 & +27.2 & +35.0 \\
Cohen's $d$ & 0.77 & 0.54 & 0.49 & 0.68 & 0.49 & 0.51 \\
Win rate & 83\% & 84\% & 78\% & 82\% & 77\% & 79\% \\
\bottomrule
\end{tabular}
}

\caption{Mean per-module branch coverage ($\pm$ standard error) for all 3 models. CovQValue vs.\ Greedy comparisons: $p < 0.0001$ (paired $t$-test). $d$: Cohen's $d$ effect size.}
\label{tab:main}
\end{table*}

\begin{figure}[t]
    \centering
    \begin{subfigure}[t]{0.48\linewidth}
        \centering
        \includegraphics[width=\linewidth]{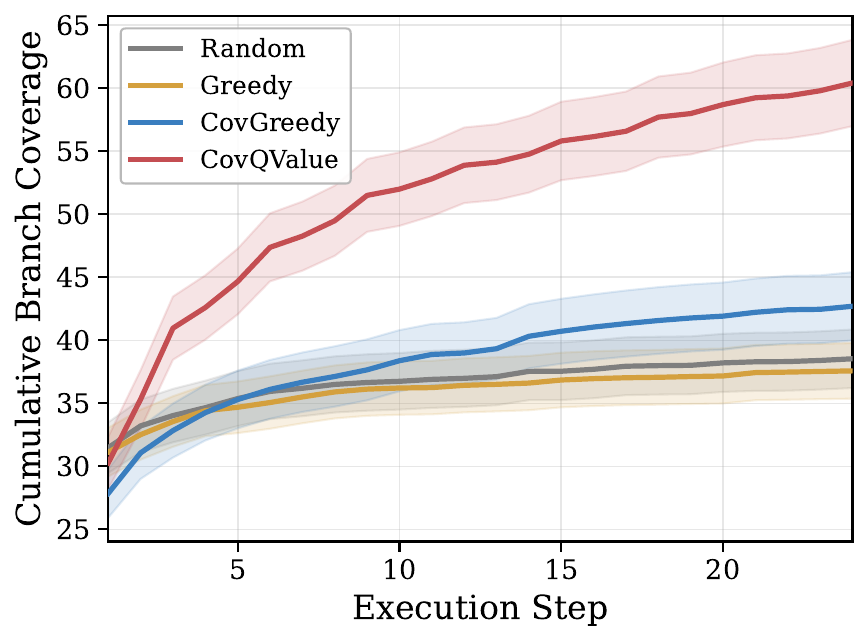}
        \caption{RepoExploreBench}
        \label{fig:exploration_reb}
    \end{subfigure}
    \hspace{10pt}
    \begin{subfigure}[t]{0.48\linewidth}
        \centering
        \includegraphics[width=\linewidth]{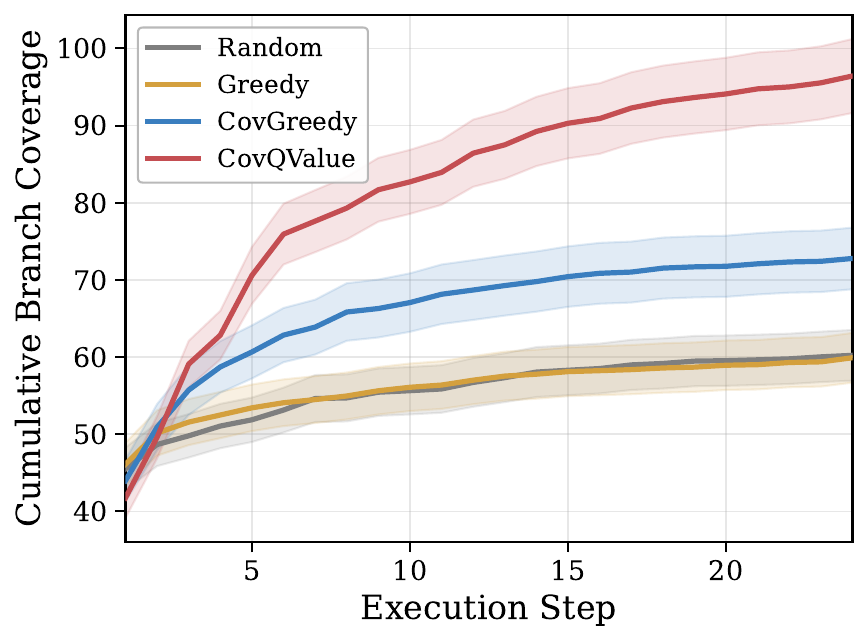}
        \caption{TestGenEval Lite}
        \label{fig:exploration_tge}
    \end{subfigure}
    \caption{Cumulative branch coverage over execution steps, averaged across three models (Gemini 3 Flash, GPT-5.4 Mini, Mistral Large). CovQValue (ours) continues discovering branches while Random and Greedy stall earlier. Shaded region indicates $\pm$1 SE.}
    \label{fig:exploration_curves}
\end{figure}

\begin{figure}[t]
    \vspace{-12pt}
    \centering
    \includegraphics[width=0.95\linewidth]{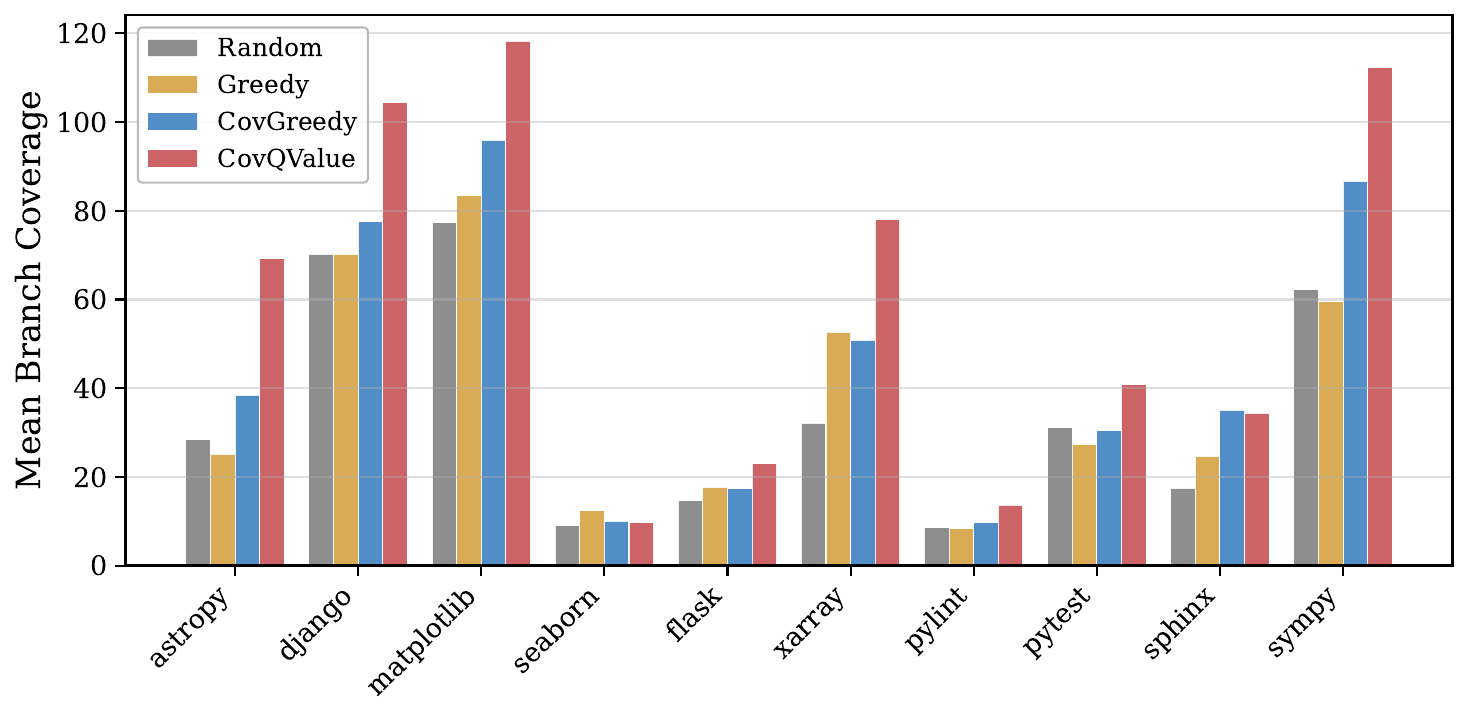}
    \caption{Mean branch coverage by repository on TestGenEval Lite, averaged across three models. CovQValue outperforms other methods on every repository, with the largest gains over Greedy on sympy (+55\%), matplotlib (+52\%), and astropy (+55\%).}
    \label{fig:per_repo_tge}
    \vspace{-10pt}
\end{figure}
                                         
\paragraph{CovQValue outperforms baselines in all models and benchmarks} Table \ref{tab:main} shows branch coverage in all settings. CovQValue achieves the highest coverage in every model, with improvements over Greedy ranging from +16.8 to +47.4 branches. The method wins on 77--84\% of individual samples and shows that it generalizes over the different models families evaluated. CovGreedy, which adds coverage map feedback but no planning, shows a moderate improvement, but it is not as consistent. Greedy method does not perform better than Random and confirms that LLM-based selection without observational feedback cannot predict which test will discover new branches from code inspection alone.

\paragraph{Coverage grows with exploration budget} Figure~\ref{fig:exploration_curves} presents the cumulative branch coverage over execution steps, averaged for all models (Individual plots available in Appendix \ref{app:evolution_per_model}). We observe that Random and Greedy strategies plateau earlier, around 5 steps, after which they only generate variations of the same tests without making fundamental modifications. CovGreedy climbs steadily as the coverage map shows untested code areas. Finally, CovQValue rises fast with steep increases when discovering a corridor. The increasing gap over time between CovQValue and the rest is consistent with the cumulative nature of information gain, where each discovery updates the posterior and enables the next.

\begin{wrapfigure}{r}{0.5\textwidth}
    \vspace{-15pt}
    \centering
    \includegraphics[width=\linewidth]{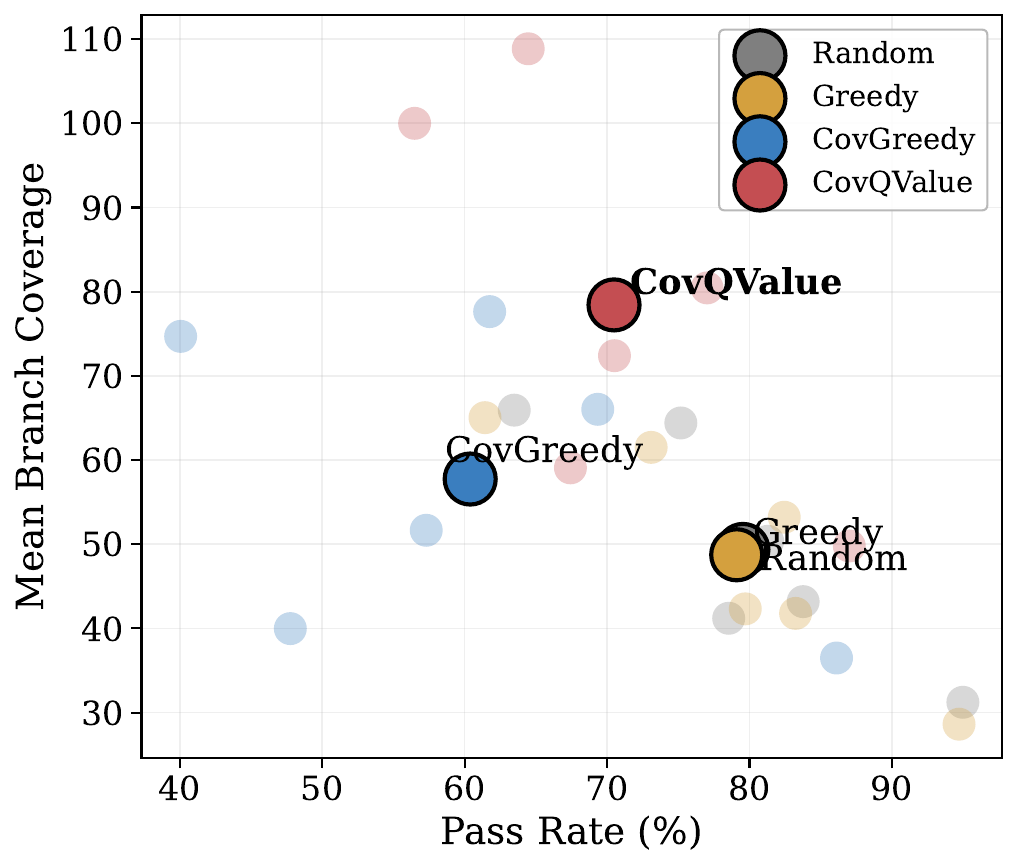}
    \caption{Trade-off between test pass rate and branch coverage. Large dots represent the mean across models, while individual models and benchmark results are represent with transparent dots. Coverage-map strategies discover more branches by generating larger tests at a small cost in pass rate. }
    \label{fig:pass_rate_vs_coverage}
    \vspace{-20pt}
\end{wrapfigure}

\paragraph{Consistent results across repositories} Figure \ref{fig:per_repo_tge} shows coverage split by repository on TestGenEval Lite, averaged for all models. CovQValue shows the best performance for all methods on every repository. We see larger gains on repositories such as sympy (+55\%), matplotlib (+52\%), and astropy (+55\%), which are among the most complex repositories in the benchmark. Even on repositories where all strategies achieve low coverage, CovQValue matches or exceeds the results from other methods. Results for ExploreBench can be found in Appendix \ref{app:per-repo-data} and show the same pattern where CovQValue leads on all repositories.

\paragraph{Coverage gains come from deeper exploration, not safer tests} Figure \ref{fig:pass_rate_vs_coverage} plots the relationship between test pass rate and branch coverage. Random and Greedy cluster at high pass rates (75--95\%) but low coverage. This means they generate safe tests that revisit the same code. Meanwhile, CovQValue and CovGreedy achieve lower pass rates (57--71\%), meaning the coverage map pushes the LLM to write more ambitious tests to target unexplored code paths. These tests tend to fail more often, but they cover more branches. From the plot, we infer that CovQValue achieves the best tradeoff, where it reaches the highest coverage at the expense of pass rate. Full line coverage and pass rate results (available in Appendix~\ref{app:lines}) confirm these findings are consistent under both metrics.

\subsection{Ablation Studies}

\begin{figure*}[t]
    \centering
    \begin{minipage}[c]{0.30\textwidth}
        \centering
        \includegraphics[width=\linewidth]{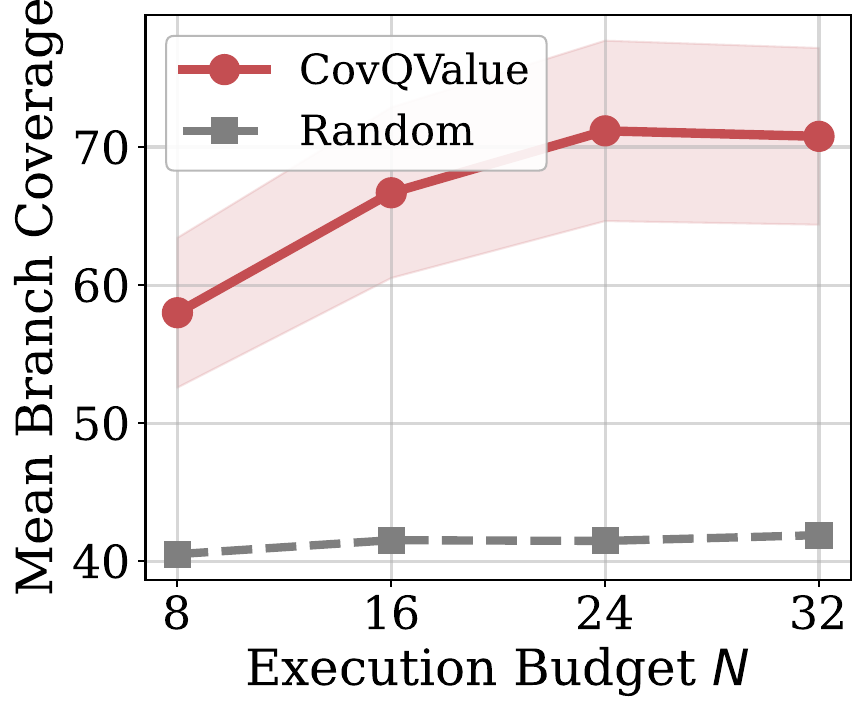}
        \subcaption{Execution budget $N$.}
        \label{fig:ablation_budget}
    \end{minipage}
    \hfill
    \begin{minipage}[c]{0.30\textwidth}
        \centering
        \includegraphics[width=\linewidth]{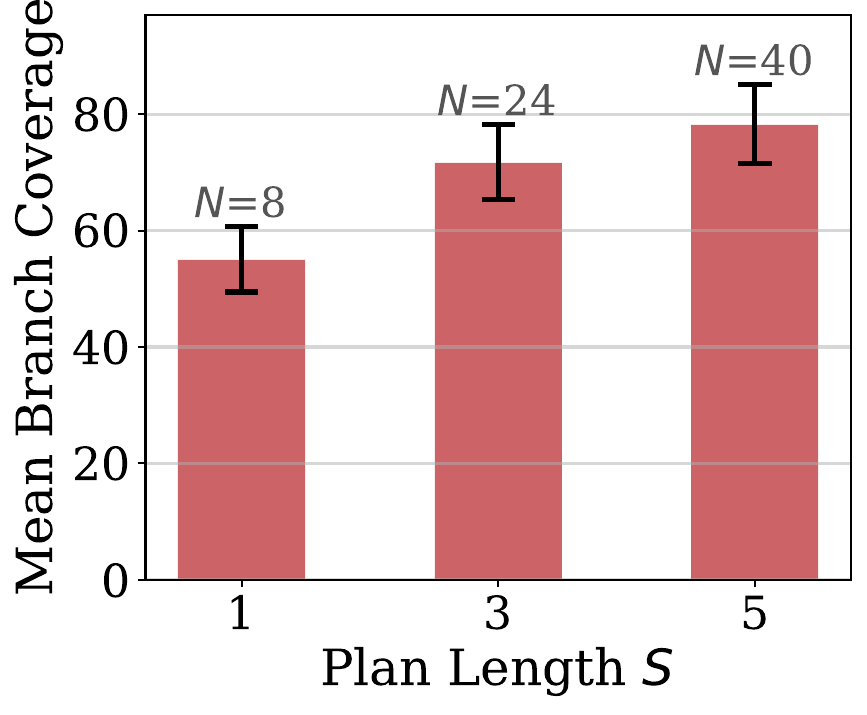}
        \subcaption{Plan length $S$.}
        \label{fig:ablation_plan_length}
    \end{minipage}
    \hfill
    \begin{minipage}[c]{0.33\textwidth}
        \centering
        \small
        \setlength{\tabcolsep}{3pt}
        \begin{tabular}{lc}
            \toprule
            \textbf{Setting} & \textbf{Coverage} \\
            \midrule
            $\gamma = 0.0$ & 68.6 \\
            $\gamma = \textbf{0.5}$ & \textbf{71.1} \\
            $\gamma = 1.0$ & 70.0 \\
            \midrule
            \midrule
            Cov.\ map & 49.3 {\scriptsize$\pm$5.4} \\
            \quad + Div.\ hints & 67.9 {\scriptsize$\pm$6.1} \\
            \quad + Q-val.\ scoring & 72.4 {\scriptsize$\pm$6.5} \\
            \bottomrule
        \end{tabular}
        \subcaption{Hyperparams. \& components.}
        \label{tab:ablations}
    \end{minipage}
    \caption{Ablation studies on RepoExploreBench (93 targets, Gemini 3 Flash). (a)~CovQValue scales steadily with budget. (b)~Longer plans yield higher coverage, confirming the benefit of multi-step lookahead. (c)~Top: discount factor sensitivity (default in bold). Bottom: component contributions.}
    \label{fig:ablations}
\end{figure*}

We run ablation studies for the key components of CovQValue on RepoBench with Gemini 3 Flash to understand what is driving the improvement. Results are summarized in Table~\ref{tab:ablations} and Figure~\ref{fig:ablations}.

\textbullet\hspace{0.5em} \textbf{Execution budget} Figure~\ref{fig:ablation_budget} shows coverage as a function of the execution budget $N$. CovQValue scales from 58.0 ($N{=}8$) to 71.2 ($N{=}24$) branches before stalling, while Random stays more or less flat at $\sim$41. This confirms that the coverage map is a key factor enabling the discovery process.

\textbullet\hspace{0.5em} \textbf{Plan length} Figure~\ref{fig:ablation_plan_length} shows coverage as a function of plan length $S$, with the number of planning rounds fixed at 8.  Total executions increase with $S$ ($8 \times S$), but Random at $N{=}40$ reaches only $\sim$43 branches (Figure~\ref{fig:ablation_budget}) while CovQValue with $S{=}5$ at the same budget reaches 79, confirming that multi-step plans contribute beyond additional executions.

\textbullet\hspace{0.5em} \textbf{Discount factor ($\gamma$)} The discount factor controls how much the Q-value scorer weighs future reachability relative to immediate branch discovery. Coverage is highest at $\gamma{=}0.5$ (71.1) and lowest at $\gamma{=}0$ (68.6). The gap is modest but consistent, which tells us the future value provides a useful signal for plan selection. Designing better future-value estimators is an area of improvement.

 \textbullet\hspace{0.5em} \textbf{Component contributions} Table~\ref{tab:ablations} (bottom) decomposes the method over the coverage-map baseline. Adding generation diversity alone raises coverage by +18.6 points, while adding Q-value scoring alone raises it by +20.5. Both components independently close most of the gap to the full method (+18.6 and +20.5 over the baseline, vs.\ +23.1 for the full method), suggesting they partially address the same bottleneck with modest complementary improvements when combined.

\subsection{Case Study}

\begin{figure}[h]
  \centering
  \includegraphics[width=\textwidth]{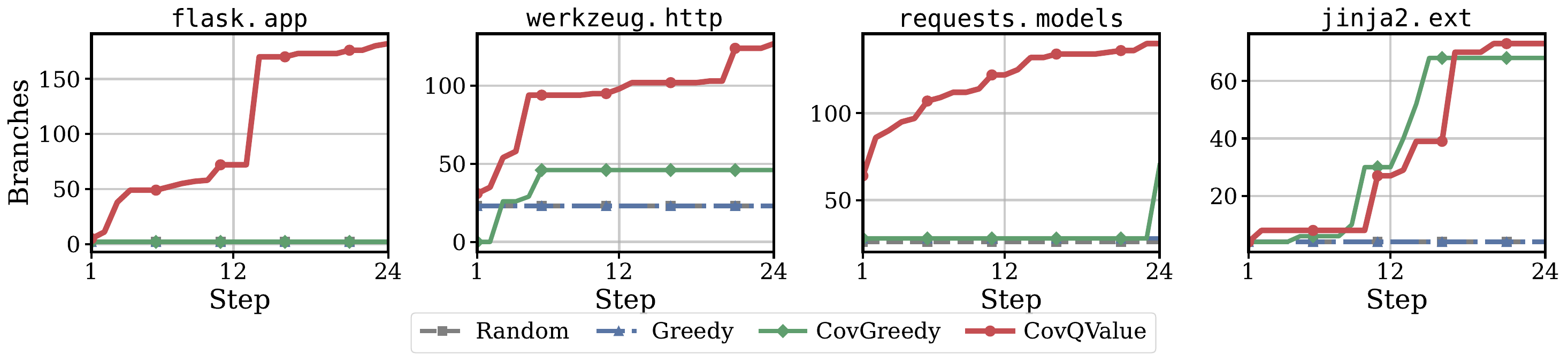}
  \caption{Corridor navigation on four RepoExploreBench modules (Gemini 3 Flash). It showcases how Random or Greedy remain flat while CovQValue traverses setup corridors to reach deep branches as the number of steps increase.}
  \label{fig:case_studies}
  \vspace{-10pt}
\end{figure}

To illustrate how CovQValue navigates corridor structure, Figure~\ref{fig:case_studies} shows the step-by-step branch coverage for four modules from RepoExploreBench with Gemini 3 Flash. In \texttt{flask.app}, all three baselines remain stuck at 2 branches for all 24 steps. The Flask application factory requires a specific initialization sequence that none of them discover. CovQValue breaks through immediately and reaches 182 branches. In \texttt{werkzeug.http} and \texttt{requests.models}, Random and Greedy are flat while CovQValue climbs steadily, with CovGreedy achieving partial but limited progress. In \texttt{jinja2.ext}, CovQValue is flat in the initial steps before getting into the extension registration corridor. Furthermore, we go over a concrete example in Appendix \ref{app:case_study_scripts}. These examples help to show how the method works and how Q-value scorer recognizes their future reachability of certain steps.

\section{Conclusions} 

In this work, we formalized LLM-base test generation as Bayesian exploration, using the coverage map as the sufficient statistic summarizing exploration state and  Q-value scoring to select among diverse candidate test plans. To evaluate our method, we introduced the benchmark, \emph{RepoExploreBench}, designed for evaluating iterative exploration-based test generation. We evaluated our method, \emph{CovQValue}, on RepoExploreBench and the public benchmark TestGenEval Lite. Our results show that CovQValue achieves higher branch coverage than greedy generation with consistent improvements across multiple LLMs. Our analysis shows that the key to closing the loop between feedback and generation is the coverage feedback, without it the LLM-based selection is practically random. This aligns with the theoretical framework where there is no information gain to optimize without posterior updates. The framework can naturally include richer value estimators (e.g. execution-based scoring, multi-step rollout) and can be an area of future research. We believe that the frameworks can also be applied to other settings where LLM agents must discover structure through interaction with an unknown environment and can become a general-purpose exploration module for agentic systems operating under partial observability.

\section*{Ethics Statement}

Automated test generation can contribute to software reliability by increasing the coverage and precision of code evaluation. As more AI-generated code is deployed in the real world, so does the need for automatically generated tests. In this paper, we show methods that generate tests automatically without human review in the loop, and practitioners should treat generated tests with the same caution as other automatically generated tests. 

Our framework gives the LLM greater autonomy in deciding what to explore, guided by its own metrics rather than explicit human instructions. In our experimental setting, exploration is naturally bounded by the program under test and the coverage map. In more open-ended real-world deployments, where an LLM agent interacts with live systems, APIs, or external services, such natural boundaries may not exist, and additional guardrails would be necessary to prevent unintended or harmful exploratory actions.

We acknowledge that further scaling these agents to other domains demands careful consideration of what constitutes safe exploratory actions. Future work should investigate guardrails that constrain the space of exploration, as well as the domains to which such methods could extend.

\section*{Reproducibility Statement}
\vspace{-3pt}

Link to the source code: \url{https://github.com/amayuelas/qcurious-tester}

The paper commits to disclose and make publicly available the code, benchmark definitions, experiment configuration and results. RepoExploreBench, is built on top of open-source repositories, its targets and the Docker image for coverage measurement are included in the repository. TestGenEval Lite uses publicly available SWE-bench Docker images. 

All three LLMs used for the experiments are accessible through public APIs. Additionally, one of the models, Mistral Large 3, is available with open-weight for local use on Huggingface. Hyperparameters are fixed in the main experiments and reported in Section~\ref{Sec:Experiments}.

In compliance with the guidelines, LLMs are the subjects of this research and were used to generate all the experimental data. LLMs were used to assist with editing and proofreading of this paper. Additionally, AI pair-programming and agentic tools were used as a coding assistant during the implementation and paper preparation. 

\bibliography{colm2026_conference}
\bibliographystyle{colm2026_conference}

\appendix

\newpage


\section{RepoExploreBench Benchmark Creation}
\label{app:RepoExploreBench}

RepoExploreBench is our benchmark designed to evaluate exploration-based test generation on real-world Python code available online with corridor structure.

\paragraph{Repository selection.} We selected 9 packages from the top 150 most-downloaded (as of March 2026) PyPI packages (30-day rolling window) using three criteria: (1) pure Python source with $\geq$5K lines, (2) not C extensions, build tools, or type stubs, and (3) exhibit corridor structure such as deep import chains, class hierarchies, and configuration requirements.

\paragraph{Module selection.} From each package, we selected approximately 10 modules using the following criteria:

\begin{enumerate}[nosep]
    \item Minimum 200 source lines (enough branches for meaningful exploration)
    \item Larger modules preferred (more paths to discover)
    \item Functional diversity within each repo (avoid selecting similar utilities)
    \item Each module must expose callable classes or functions, not just constants
    \item Exclude test-only modules (except testing utilities such as \texttt{click.testing})
    \item Exclude private modules (except httpx, which uses \texttt{\_} prefix by convention)
    \item Exclude vendored code and legacy compatibility layers
\end{enumerate}                                                                                                             

Module candidates were identified by enumerating all submodules in each package and verifying line counts inside. Each package contains many more modules meeting the minimum size criterion; we sample approximately 10 per package for a manageable evaluation set, but the benchmark can be scaled without additional infrastructure. This yields 93 targets totaling 77,242 lines of code. Table~\ref{tab:repoexplorebench_composition} summarizes the benchmark composition.

\begin{table}[h]
\centering
\small
\resizebox{\textwidth}{!}{%
\begin{tabular}{l l r r l}
\toprule
\textbf{Package} & \textbf{Domain} & \textbf{Mod.} & \textbf{Lines} & \textbf{Corridor Structure} \\
\midrule
click      & CLI framework       & 10 & 8,293  & Command setup → option parsing → type coercion \\
flask      & Web framework       & 10 & 5,369  & App creation → config → routing → request context \\
httpx      & Async HTTP client   & 10 & 6,925  & Client config → auth → transport → URL parsing \\
jinja2     & Template engine     & 12 & 12,751 & Env setup → lexer → parser → AST → compiler \\
pydantic   & Data validation     & 10 & 11,727 & Model definition → schema → validation → serialization \\
requests   & HTTP client         & 6  & 4,375  & Session setup → auth → adapters → connection pool \\
rich       & Terminal rendering  & 12 & 12,792 & Console setup → style parsing → segment → rendering \\
starlette  & ASGI framework      & 11 & 4,625  & App → routing → middleware → request/response \\
werkzeug   & WSGI utilities      & 12 & 10,385 & Routing setup → request parsing → response building \\
\midrule
\textbf{Total} & & \textbf{93} & \textbf{77,242} & \\
\bottomrule
\end{tabular}}%
\caption{RepoExploreBench composition. Packages selected from top 150 PyPI downloads (as of March 2026). Corridor structure describes the sequential dependencies that gate access
 to deep branches.}
\label{tab:repoexplorebench_composition}
\end{table}

\paragraph{Execution environment.} All modules execute inside a single Docker image with all 9 packages pre-installed. Branch coverage is measured using Python's standard branch coverage tooling.   


\section{Algorithm}
\label{app:Algorithm}

Our method, which we call \textbf{CovQValue}, operates as follows. Given a target source file and an execution budget of N steps:


\begin{algorithm}[!h]
\caption{\textsc{CovQValue}: Curiosity-Driven Test Plan Selection}
\label{alg:covqvalue}
\begin{algorithmic}[1]
\REQUIRE Source file $P$, execution budget $N$, number of plans $K$, discount $\gamma$, steps per plan $S$
\ENSURE Coverage map $\mathcal{C}$

\STATE $\mathcal{C} \leftarrow \emptyset$ \COMMENT{Coverage map (posterior)}
\STATE $h \leftarrow \emptyset$ \COMMENT{Test history}

\WHILE{executions remaining $> 0$}

    \STATE \textit{// Generate $K$ candidate plans in parallel}
    \FOR{$k = 1$ \TO $K$}
        \STATE $\pi_k \leftarrow \textsc{LLM.GeneratePlan}(P, \mathcal{C}, h, \text{hint}_k)$
        \COMMENT{Each $\pi_k = (t_1^k, \ldots, t_S^k)$}
    \ENDFOR

    \STATE \textit{// Score each plan by curiosity Q-value}
    \FOR{$k = 1$ \TO $K$}
        \STATE $\bar{g}_k \leftarrow \textsc{EstimateImmediateGain}(\pi_k, \mathcal{C})$
        \STATE $\hat{v}_k \leftarrow \textsc{EstimateFutureReachability}(\pi_k, \mathcal{C})$
        \STATE $Q_k \leftarrow \bar{g}_k + \gamma \cdot \hat{v}_k$
    \ENDFOR

    \STATE \textit{// Select and execute best plan}
    \STATE $k^* \leftarrow \arg\max_k Q_k$
    \FOR{$s = 1$ \TO $S$}
        \STATE $o_s \leftarrow \textsc{Execute}(P, t_s^{k^*})$
        \STATE $\mathcal{C} \leftarrow \mathcal{C} \cup \textsc{Branches}(o_s)$
        \STATE $h \leftarrow h \cdot (t_s^{k^*}, o_s)$
    \ENDFOR

\ENDWHILE

\RETURN $\mathcal{C}$
\end{algorithmic}
\end{algorithm}

Each plan consists of $S$ test scripts, designed as a coherent sequence. For example, the first script may set up fixtures, the second may exercise a specific code path, and the third may probe edge cases exposed by the previous steps. The diversity hints encourage different exploration strategies across the K plans (e.g., "focus on untested exception paths," "target the deepest nested branches," "explore import-chain dependencies").


\section{Line Coverage and Pass Rates}
\label{app:lines}

Table~\ref{tab:lines} reports \textbf{line coverage} and \textbf{test pass rates} for all strategies and models on both benchmarks. The trends mirror the branch coverage results. CovQValue achieves the highest line coverage in every model and benchmark, while Random and Greedy plateau at lower values. Pass rates for coverage-aware strategies (CovGreedy and CovQValue) are lower than for Random and Greedy, consistent with the observation in Section~\ref{Sec:Results} that guiding generation toward uncovered code produces more ambitious tests that fail more often but discover more code when they succeed.

\begin{table}[h]
\centering
\begin{tabular}{l|cc|cc|cc}
\toprule
 & \multicolumn{2}{c|}{Gemini 3 Flash} & \multicolumn{2}{c|}{GPT-5.4 Mini} & \multicolumn{2}{c}{Mistral Large} \\
\textbf{Strategy} & Lines & Pass\% & Lines & Pass\% & Lines & Pass\% \\
\midrule
\multicolumn{7}{l}{\textit{RepoExploreBench}} \\
\quad Random & 146 & 84\% & 113 & 95\% & 142 & 79\% \\
\quad Greedy & 142 & 83\% & 109 & 95\% & 144 & 80\% \\
\quad CovGreedy & 157 & 57\% & 123 & 86\% & 129 & 48\% \\
\quad CovQValue & \textbf{197} & 71\% & \textbf{151} & 87\% & \textbf{171} & 67\% \\
\midrule
\multicolumn{7}{l}{\textit{TestGenEval Lite}} \\
\quad Random & 149 & 75\% & 135 & 81\% & 148 & 63\% \\
\quad Greedy & 149 & 73\% & 141 & 82\% & 148 & 61\% \\
\quad CovGreedy & 161 & 62\% & 151 & 69\% & 157 & 40\% \\
\quad CovQValue & \textbf{193} & 64\% & \textbf{164} & 77\% & \textbf{180} & 56\% \\
\bottomrule
\end{tabular}
\caption{Line coverage and pass rate. Results consistent with branch coverage.}
\label{tab:lines}
\end{table}


\section{Per-repo Data}
\label{app:per-repo-data}

\begin{figure}[h]
    \centering
    \includegraphics[width=\linewidth]{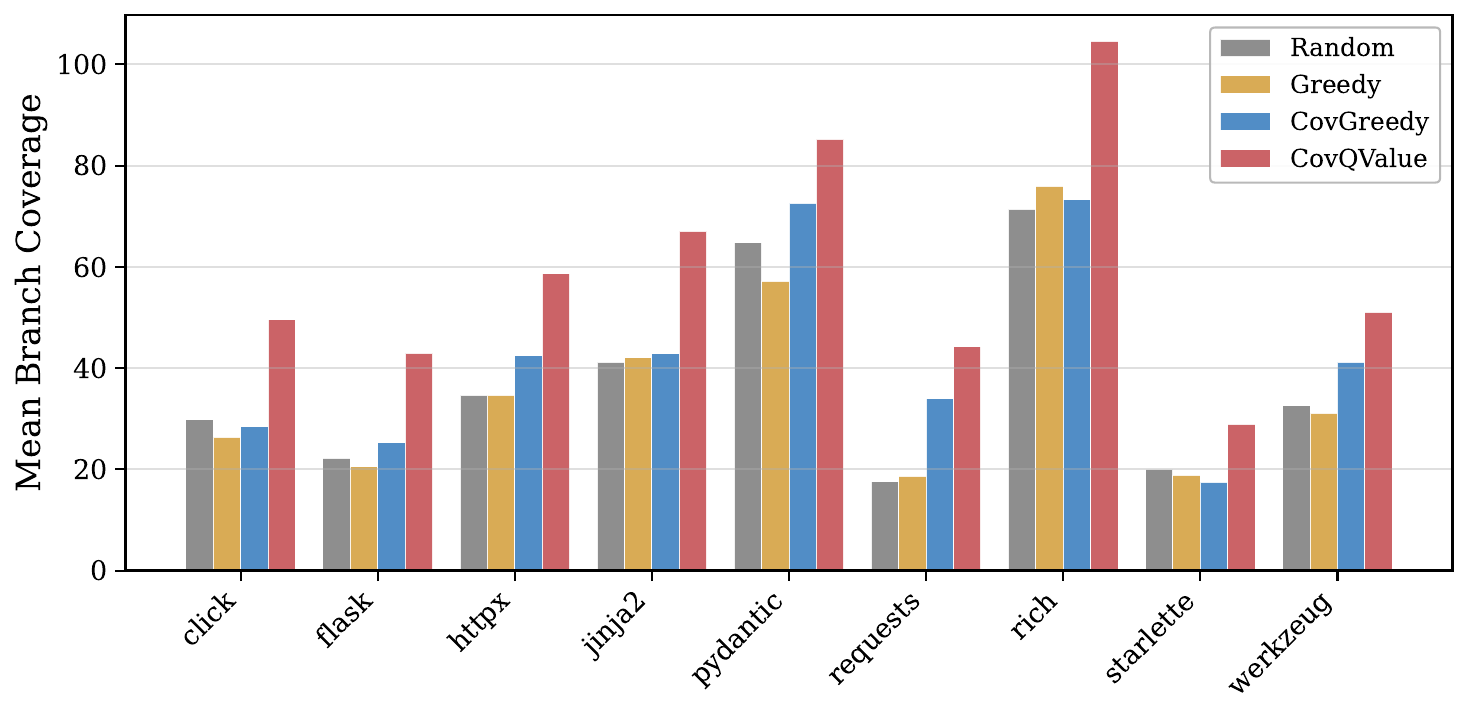}
    \caption{Mean branch coverage by repository on RepoExploreBench,
  averaged for all models. CovQValue outperforms all baselines
  on every repository.}
    \label{fig:per_repo_reb}
\end{figure}

Here, we provide per-repository breakdowns of branch coverage for both benchmarks. Figure~\ref{fig:per_repo_reb} visualizes the mean coverage on RepoExploreBench averaged for all models. Tables~\ref{tab:perrepo_testgeneval} and~\ref{tab:perrepo_repoexplorebench} give the full numbers for TestGenEval Lite and RepoExploreBench in every model. CovQValue achieves the highest coverage on every repository in both benchmarks. The largest absolute gains appear on repositories with deep branching logic (e.g., \texttt{sympy}, \texttt{rich}, \texttt{matplotlib}), where multi-step planning can chain discoveries for dependent paths.

\begin{table}[h]
\centering
\setlength{\tabcolsep}{3pt}
\begin{tabular}{l cccc cccc cccc}
\toprule
& \multicolumn{4}{c}{Gemini Flash}& \multicolumn{4}{c}{GPT-5.4 Mini}& \multicolumn{4}{c}{Mistral Large 3} \\
\cmidrule(lr){2-5} \cmidrule(lr){6-9} \cmidrule(lr){10-13}
Repo & Rnd & Gdy & CG & CQ & Rnd & Gdy & CG & CQ & Rnd & Gdy & CG & CQ \\
\midrule
astropy & 37 & 23 & 49 & \textbf{80} & 13 & 18 & 19 & \textbf{51} & 35 & 35 & 47 & \textbf{76} \\
django & 69 & 71 & 83 & \textbf{114} & 61 & 65 & 73 & \textbf{90} & 81 & 74 & 76 & \textbf{109} \\
matplotlib & 93 & 93 & 101 & \textbf{133} & 53 & 68 & 81 & \textbf{103} & 86 & 89 & 106 & \textbf{119} \\
seaborn & 10 & 20 & 10 & \textbf{10} & 9 & 9 & 10 & \textbf{10} & 8 & 8 & 9 & \textbf{9} \\
flask & 10 & 18 & 23 & \textbf{23} & 19 & 17 & 21 & \textbf{25} & 15 & 18 & 8 & \textbf{21} \\
xarray & 35 & 35 & 52 & \textbf{81} & 41 & 42 & 48 & \textbf{54} & 21 & 81 & 53 & \textbf{100} \\
pylint & 7 & 7 & 9 & \textbf{20} & 10 & 10 & 10 & \textbf{10} & 9 & 8 & 10 & \textbf{10} \\
pytest & 32 & 28 & 38 & \textbf{50} & 23 & 19 & 19 & \textbf{22} & 38 & 36 & 34 & \textbf{50} \\
sphinx & 17 & 30 & 37 & \textbf{37} & 10 & 18 & 34 & \textbf{32} & 26 & 26 & 34 & \textbf{34} \\
sympy & 75 & 63 & 90 & \textbf{132} & 50 & 52 & 78 & \textbf{92} & 62 & 64 & 92 & \textbf{112} \\
\midrule
Mean & 64.4 & 61.5 & 77.6 & \textbf{108.9} & 50.4 & 53.2 & 66.0 & \textbf{80.5} & 65.9 & 65.0 & 74.7 & \textbf{100.0} \\
\bottomrule
\end{tabular}
\caption{Per-repository branch coverage on TestGenEval Lite by model. Each cell averages across modules in that repo. Mean row shows the per-module average. Best per group in bold.}
\label{tab:perrepo_testgeneval}
\end{table}

\begin{table}[h]
\centering
\setlength{\tabcolsep}{3pt}
\begin{tabular}{l cccc cccc cccc}
\toprule
& \multicolumn{4}{c}{Gemini Flash}& \multicolumn{4}{c}{GPT-5.4 Mini}& \multicolumn{4}{c}{Mistral Large 3} \\
\cmidrule(lr){2-5} \cmidrule(lr){6-9} \cmidrule(lr){10-13}
Repo & Rnd & Gdy & CG & CQ & Rnd & Gdy & CG & CQ & Rnd & Gdy & CG & CQ \\
\midrule
click & 38 & 28 & 32 & \textbf{68} & 23 & 24 & 27 & \textbf{29} & 29 & 27 & 26 & \textbf{51} \\
flask & 19 & 24 & 31 & \textbf{53} & 28 & 18 & 28 & \textbf{39} & 20 & 20 & 17 & \textbf{37} \\
httpx & 37 & 38 & 45 & \textbf{67} & 33 & 33 & 39 & \textbf{54} & 35 & 33 & 44 & \textbf{54} \\
jinja2 & 51 & 46 & 57 & \textbf{84} & 29 & 30 & 32 & \textbf{56} & 43 & 50 & 40 & \textbf{61} \\
pydantic & 70 & 58 & 83 & \textbf{88} & 52 & 48 & 58 & \textbf{80} & 72 & 66 & 76 & \textbf{87} \\
requests & 17 & 20 & 46 & \textbf{66} & 21 & 19 & 31 & \textbf{31} & 14 & 17 & 25 & \textbf{36} \\
rich & 88 & 94 & 103 & \textbf{128} & 42 & 35 & 45 & \textbf{81} & 84 & 99 & 72 & \textbf{105} \\
starlette & 23 & 23 & 19 & \textbf{34} & 18 & 16 & 20 & \textbf{21} & 20 & 18 & 13 & \textbf{32} \\
werkzeug & 31 & 30 & 40 & \textbf{54} & 30 & 30 & 46 & \textbf{46} & 37 & 33 & 37 & \textbf{53} \\
\midrule
Mean & 43.2 & 41.8 & 51.7 & \textbf{72.4} & 31.2 & 28.6 & 36.5 & \textbf{49.8} & 41.2 & 42.3 & 40.0 & \textbf{59.1} \\
\bottomrule
\end{tabular}
\caption{Per-repository branch coverage on RepoExploreBench by model. Each cell averages across modules in that repo. Mean row shows the per-module average Best per group in bold.}
\label{tab:perrepo_repoexplorebench}
\end{table}


\newpage
\section{Individual Cumulative Branch Coverage per model}
\label{app:evolution_per_model}

For informative purposes, we also show the cumulative branch coverage evolution presented for each model individually, instead of the average as presented in Figure \ref{fig:exploration_curves}

\begin{figure}[h]
    \centering
    \begin{subfigure}[b]{0.3\textwidth}
        \centering
        \includegraphics[width=\textwidth]{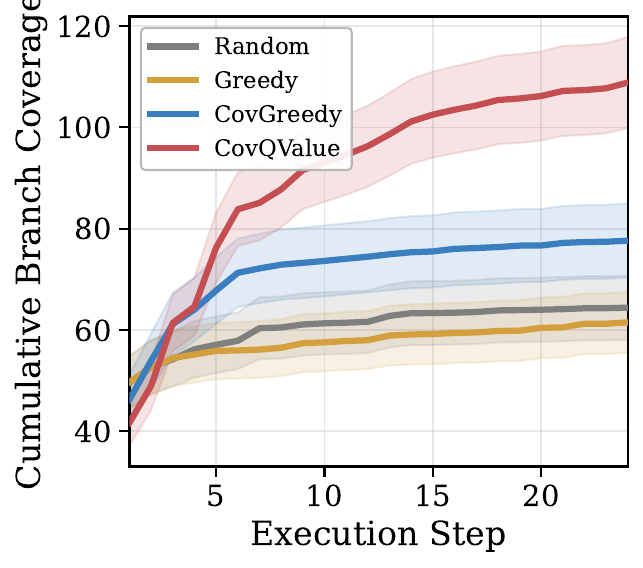}
        \caption{Gemini 3 Flash}
        \label{fig:exploration_curves_tge_gemini}
    \end{subfigure}
    \hfill
    \begin{subfigure}[b]{0.3\textwidth}
        \centering
        \includegraphics[width=\textwidth]{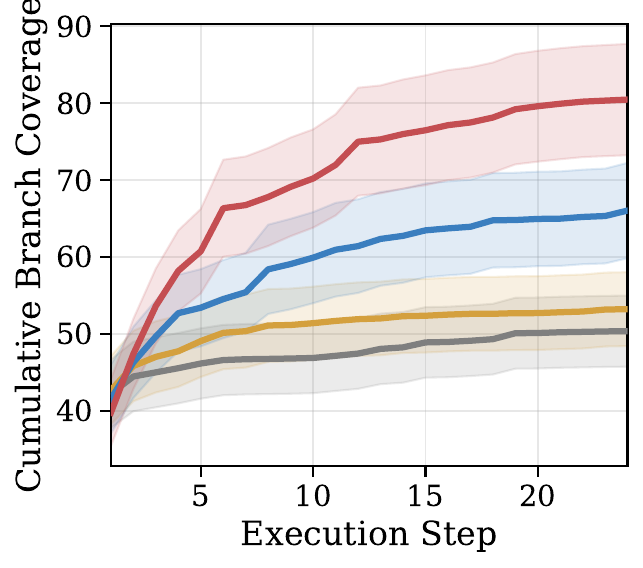}
        \caption{GPT-5.4 mini}
        \label{fig:exploration_curves_tge_gpt54mini}
    \end{subfigure}
    \hfill
    \begin{subfigure}[b]{0.3\textwidth}
        \centering
        \includegraphics[width=\textwidth]{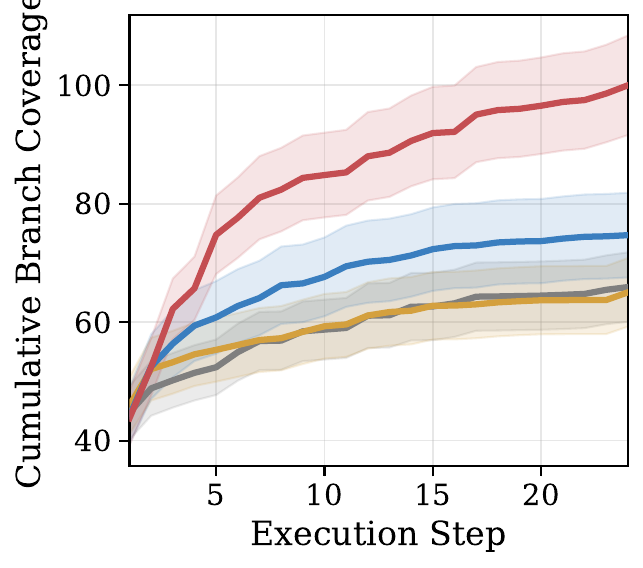}
        \caption{Mistral Large 3}
        \label{fig:exploration_curves_tge_mistral}
    \end{subfigure}
    \captionof{figure}{Cumulative branch coverage over execution steps on TestGenEval Lite plotted individually for each model.}
    \label{fig:subfigures}
\end{figure}

\begin{figure}[h]
    \centering
    \begin{subfigure}[b]{0.3\textwidth}
        \centering
        \includegraphics[width=\textwidth]{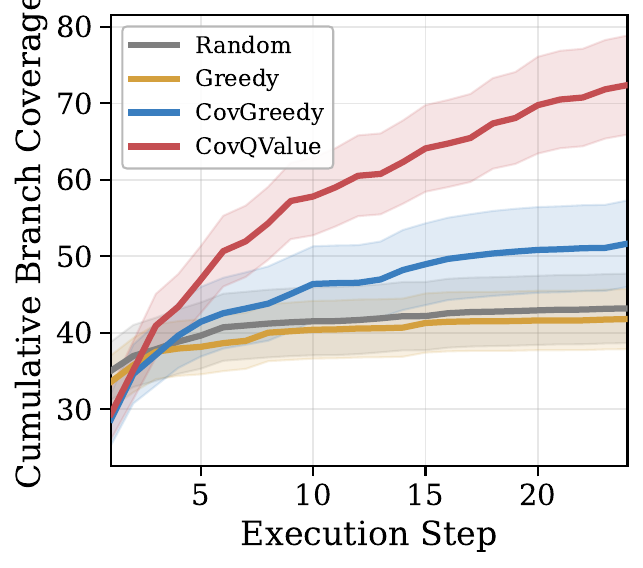}
        \caption{Gemini 3 Flash}
        \label{fig:exploration_curves_reb_gemini}
    \end{subfigure}
    \hfill
    \begin{subfigure}[b]{0.3\textwidth}
        \centering
        \includegraphics[width=\textwidth]{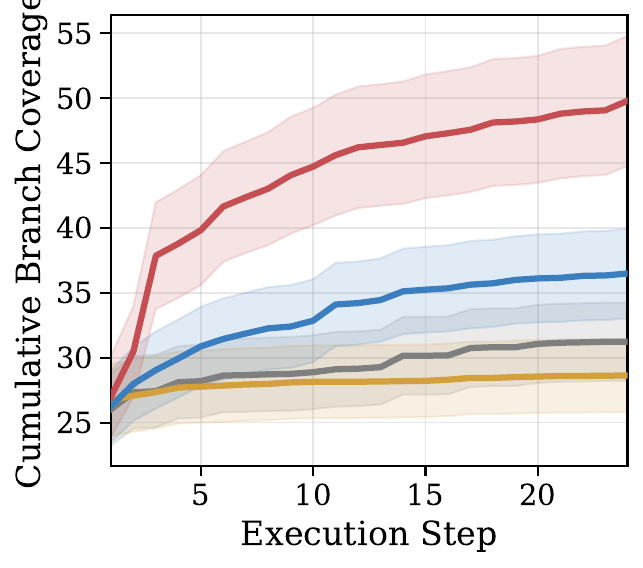}
        \caption{GPT-5.4 mini}
        \label{fig:exploration_curves_reb_gpt54mini}
    \end{subfigure}
    \hfill
    \begin{subfigure}[b]{0.3\textwidth}
        \centering
        \includegraphics[width=\textwidth]{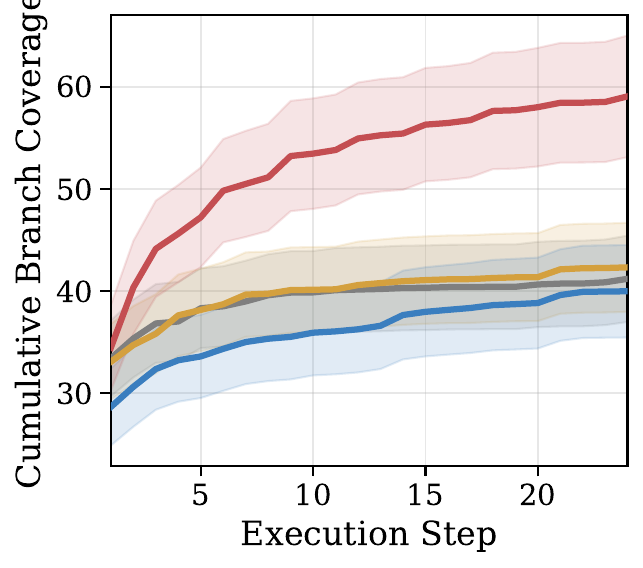}
        \caption{Mistral Large 3}
        \label{fig:exploration_curves_reb_mistral}
    \end{subfigure}
    \captionof{figure}{Cumulative branch coverage over execution steps on RepoExploreBench plotted individually for each model.}
    \label{fig:subfigures}
\end{figure}


\section{Case Study: Generated Test Scripts}
\label{app:case_study_scripts}

We examine the test scripts generated for \texttt{werkzeug.http} with Gemini 3 Flash to illustrate how coverage feedback and Q-value scoring navigate the corridor structure. The corresponding branch coverage curves are shown in Figure~\ref{fig:case_studies} (second sub-figure from the left). Greedy remains at 23 branches for all 24 steps, and CovQValue reaches 127.

\paragraph{Module structure.} \texttt{werkzeug.http} implements over 40 functions for parsing and generating HTTP headers. Surface-level functions (\texttt{parse\_date}, \texttt{quote\_header\_value}) are standalone, but the deeper parsers (\texttt{parse\_options\_header}, \texttt{parse\_cookie}) depend on internal tokenizers and quoting logic that gate most of the module's branch complexity.

\paragraph{Greedy (stuck at 23 branches).} Greedy calls three surface-level functions on step 1 and discovers 23 branches. For the remaining 23 steps, it generates minor variations of the same test, discovering zero new branches each time:

\begin{small}
\begin{verbatim}
from werkzeug.http import (HTTP_STATUS_CODES,
    quote_header_value, parse_date)

print(f"Status 200: {HTTP_STATUS_CODES.get(200)}")
print(f"Status 404: {HTTP_STATUS_CODES.get(404)}")
header = quote_header_value("foo bar")
dt = parse_date("Mon, 21 Oct 2023 20:12:00 GMT")
\end{verbatim}
\end{small}

Without coverage feedback, Greedy has no signal that the deeper parsing layer exists. The LLM selects the candidate it considers ``most likely to cover new code paths,'' but without knowing which paths have already been covered, it keeps choosing the same familiar functions.

\paragraph{CovQValue step 3 (+19 branches).} After the initial steps cover basic utilities, the coverage map shows that structured header parsing is entirely unexplored. CovQValue targets the ETag and dictionary header parsers, which require inputs with specific quoting patterns:

\begin{small}
\begin{verbatim}
from werkzeug.http import (parse_etags,
    parse_list_header, parse_dict_header)

etags = parse_etags('W/"weak", "strong"')
print(f"Strong: {etags.contains('strong')}")
print(f"Weak: {etags.contains_weak('weak')}")
list_h = parse_list_header(
    'token1, "quoted token", token2')
dict_h = parse_dict_header('a=1, b="2", c')
\end{verbatim}
\end{small}

These functions use the internal tokenizer with weak ETags (\texttt{W/} prefix), quoted strings, and key-value pairs, input patterns that the surface-level tests never needed. This opens the first corridor gate into the structured parsing layer.

\paragraph{CovQValue step 5 (+36 branches).} The largest single-step discovery targets the cookie and content-type option machinery, which sits behind the tokenizer layer just opened:

\begin{small}
\begin{verbatim}
from werkzeug.http import (parse_options_header,
    dump_options_header, parse_cookie, dump_cookie)

header = 'text/html; charset=utf-8; boundary="x"'
mimetype, params = parse_options_header(header)
cookie = dump_cookie("session", "a b c",
    max_age=3600, httponly=True, samesite="Lax")
parsed = parse_cookie('session="a b c"; other=val')
\end{verbatim}
\end{small}

\texttt{parse\_options\_header} requires a semicolon-delimited header with quoted parameter values, and \texttt{dump\_cookie} uses security attribute serialization (\texttt{httponly}, \texttt{samesite}). These functions share internal quoting and escaping logic with the parsers from step 3, so the earlier corridor traversal was necessary. The inputs would not have reached the branches behind the quoting state machine. The Q-value scorer helped discovering it.


\section{Prompts and Coverage Map}
\label{app:prompts}

In this appendix, we present the prompts used in CovQValue and show how the coverage map evolves during exploration. All prompts receive the module source code (truncated to 2500 characters), the coverage map, and recent test history.

\subsection{Coverage Map}
\label{app:coverage_map}

The coverage map is fed to the LLM as structured text. Below we show its evolution for \texttt{werkzeug.http} at three points during a CovQValue run.

\vspace{15pt}
\begin{tcolorbox}[colback=gray!5!white, colframe=gray!60!black, title=Coverage Map --- Step 1 (initial state), fonttitle=\bfseries\small]
\begin{small}
\begin{verbatim}
COVERAGE MAP:
  Branches discovered: 0
\end{verbatim}
\end{small}
\end{tcolorbox}

\vspace{15pt}
\begin{tcolorbox}[colback=gray!5!white, colframe=gray!60!black, title=Coverage Map --- After Step 3 (54 branches), fonttitle=\bfseries\small]
\begin{small}
\begin{verbatim}
COVERAGE MAP:
  Branches discovered: 54
  Last step discovered: 19 new branches
  Average per step: 18.0 branches
  Most informative tests:
    from werkzeug.http import HTTP_STATUS -> 31 branches
    from werkzeug.http import parse_etags -> 19 branches
    from werkzeug.http import parse_date, -> 4 branches
\end{verbatim}
\end{small}
\end{tcolorbox}

\vspace{15pt}
\begin{tcolorbox}[colback=red!3!white, colframe=gray!60!black, title=Coverage Map --- After Step 12 (102 branches -- stagnation), fonttitle=\bfseries\small]
\begin{small}
\begin{verbatim}
COVERAGE MAP:
  Branches discovered: 102
  Last step discovered: 0 new branches
  Average per step: 8.5 branches
  WARNING: Coverage has stagnated for 3 steps
  You may need a fundamentally different approach
  Most informative tests:
    from werkzeug.http import parse_option -> 36 branches
    from werkzeug.http import HTTP_STATUS -> 31 branches
    from werkzeug.http import parse_range_ -> 21 branches
\end{verbatim}
\end{small}
\end{tcolorbox}

In this case, the stagnation warning triggers the LLM to try fundamentally different approaches, leading to the discovery of 21 additional branches in steps 20--24.

\newpage
\subsection{Plan Generation Prompt}
\label{app:plan_prompt}

Each of the $K{=}3$ plans receives a different diversity hint. Below is the prompt for plan 1 (main functionality). The other two plans receive: ``\textit{Focus on ERROR HANDLING}'' and ``\textit{Focus on INTERACTIONS}.''

\begin{tcolorbox}[colback=blue!3!white, colframe=blue!40!black, title=Plan Generation Prompt (with diversity hint), fonttitle=\bfseries\small]
\begin{small}
\begin{verbatim}
Module: {module_name}
{source code (truncated to 2500 chars)}

{coverage map}

Previous tests:
  {test_1} -> {output} (new branches: 31)
  {test_2} -> {output} (new branches: 4)
  {test_3} -> {output} (new branches: 19)

PLAN a sequence of 3 test scripts that TOGETHER will
reach UNCOVERED code paths. Focus on the MAIN
functionality -- constructors, primary methods.

Think about what setup is needed:
Step 1: What basic setup/import is needed to reach
        deeper code?
Step 2: Building on step 1's result, what exercises
        the next layer?
Step 3: Now target the deepest uncovered branches.

For each step, write a separate test script (5-10
lines each). Import from the module and print results.

Format your response as:
### TEST 1
```python
[code]
```
### TEST 2
```python
[code]
```
### TEST 3
```python
[code]
```
\end{verbatim}
\end{small}
\end{tcolorbox}

\newpage
\subsection{Q-Value Scoring Prompt}
\label{app:qvalue_prompt}

After generating $K$ plans, each is scored by the following prompt. The LLM estimates immediate gain $\hat{g}$ and future reachability $\hat{v}$, which are combined as $Q = \hat{g} + \gamma \cdot \hat{v}$.

\begin{tcolorbox}[colback=green!3!white, colframe=green!30!black, title=Q-Value Scoring Prompt, fonttitle=\bfseries\small]
\begin{small}
\begin{verbatim}
Module: {module_name}
{source code (truncated to 2000 chars)}

{coverage map}

Consider this TEST PLAN (a sequence of 3 scripts to
execute in order):

Step 1:
```python
{script_1}
```
Step 2:
```python
{script_2}
```
Step 3:
```python
{script_3}
```

Evaluate this plan by answering TWO questions with
just numbers:

1. IMMEDIATE GAIN: How many total NEW branches (not
   yet covered) will this sequence of tests likely
   discover? Consider that later steps build on
   earlier ones. (0-50)

2. FUTURE VALUE: After executing this plan, how many
   ADDITIONAL branches become reachable by future
   tests that weren't reachable before? Consider what
   state/objects/setup the plan leaves behind. (0-50)

Format: immediate, future
Example: 15, 25
\end{verbatim}
\end{small}
\end{tcolorbox}

\noindent The LLM responds with two numbers (e.g., ``12, 20''), yielding $Q = 12 + 0.5 \times 20 = 22$. The plan with the highest $Q$ is selected for execution.

\end{document}